Paleoproteomics explained to youngsters: how did the wedding of two-dimensional electrophoresis and protein sequencing spark proteomics on: let there be light

Thierry Rabilloud 1, 2, 3*

1: CNRS UMR 5249, Laboratoire de Chimie et Biologie des Métaux, Grenoble, France

2: Université Grenoble Alpes, Laboratoire de Chimie et Biologie des Métaux, Grenoble, France

3: CEA Grenoble, iRTSV/CBM, Laboratoire de Chimie et Biologie des Métaux, Grenoble, France

*: to whom correspondence should be addressed:

Laboratoire de Chimie et Biologie des Métaux, UMR CNRS-CEA-UJF 5249, iRTSV/LCBM, CEA Grenoble, 17 rue des martyrs, F-38054 Grenoble Cedex 9, France
thierry.rabilloud@cea.fr

Abstract
Taking the opportunity of the 20th anniversary of the word « proteomics », this young adult age is a good time to remember how proteomics came from enormous progress in protein separation and protein microanalysis techniques, and from the conjugation of these advances into a high performance and streamlined working setup. However, in the history of the almost three decades that encompass the first attempts to perform large scale analysis of proteins to the current high throughput proteomics that we can enjoy now, it is also interesting to underline and to recall how difficult the first decade was. Indeed when the word was cast, the battle was already won. This recollection is mostly devoted to the almost forgotten period where proteomics was being conceived and put to birth, as this collective scientific work will never appear when searched through the keyword "proteomics".

Introduction

This year 2014 celebrates the 20th birthday of the word "proteome", which was publicly introduced in the first Siena meeting in 1994, and used for the first time in a publication shortly thereafter [1]. These two decades of development are impressive and deserve reviewing by themselves, but in this paper I would like to use the privilege of experience to explore the family tree of proteomics and to recall the history of this young science. While it is true that human languages can create words for everything and anything, these words do not stay unless they cover a sensible concept or object, and the words "proteome" and "proteomics" are no exception to this rule. The fact that these words caught on immediately, being sometimes called buzz words and even bucks words, means that they cover a reality that was already there but with no good name, and I would like to explore in this paper this time of proteomics before it was named, the paleoproteomics if I dare call it that way. So if proteomics is born in 1994, the two preceding decades of life of its parents are worth exploring, although it will drive back almost half a century ago. I will also try to explain not only the hard scientific facts, but also to replace them in the more general landscape of the molecular biosciences and of their evolution during these two decades.

1. The 70's, a glorious start

If we take this two decades period before 1994, it drives us back to 1974, and it is interesting to describe the state of molecular biosciences at that time. Protein biochemistry was first and foremost, and the mainstream was protein purification and function determination. The methods at that time were far from being miniaturized, and compared to nowadays standards, everything was up by three orders of magnitude. What we do now on microliters was done on milliliters and what we do now on milliliters was done on liters at that time, and the necessary starting material was found more often in slaughterhouses that in tiny biopsies or on small culture dishes. The basic protein separations were already at hand, both chromatography and electrophoresis, as reviewed recently for the electrophoretic separations [2], but the detection methods were of very poor sensitivity, Coomassie Blue being recently introduced for acrylamide gel staining [3, 4]. On the side of protein characterization, protein sequencing was a well-established science, due to the outstanding work of Pehr Edman [5], but even the most modern protein sequencers of that time needed milligrams of proteins to determine a protein sequence [6, 7]. Even though miserable by today's standards, this situation was much more glorious than the one of nucleic acid sequencing, which was basically non-existing at that time except for very short and abundant RNAs such as tRNAs.

In this landscape, the two most promising protein analytical electrophoretic separations, namely isoelectric focusing and SDS-PAGE, were combined to create the most powerful protein separation method, still in use today, namely two-dimensional electrophoresis. The first publication, in 1974 [8], got relatively unnoticed, and thus poorly credited. Indeed, the use of Coomassie Blue as a detection tool produced relatively poor maps in terms of number of spots. However, the next publication in 1975 [9], using radioactive labelling, was much more spectacular and got immediate attention. Yes it was possible to visualize, separate and quantify at the same time hundreds of proteins. This paper drove a lot of enthusiasm immediately (1400 citations in the first 5 years) and some of the pioneers of proteomics started this type of large-scale protein analysis in the late 70s [10-12].

It must be realized, however, that this technique was much more an art than an everyday laboratory routine, and even worse was NEPHGE, devised to analyze the basic proteins that escape classical isoelectric focusing [13]. The isoelectric focusing gels were particularly nightmarish, as they were cast in thin, low percentage polyacrylamide rod gels. The closest description that can be made would be an overcooked rice noodle. This gives a good impression of the texture and strength of these IEF gels, which were to be loaded on top of the stronger SDS gels without any bubble. Needless to say, deformations and breakages were numerous, and it was quite an ordeal to end with a small series of comparable gels. Beyond these day to day problems, long term reproducibility was basically non existent, as the pH gradient was generated by carrier ampholytes, i.e. a modern version of a witch broth concocted by awful batch syntheses [14]. Even though different batches produced the same final pH gradient in its span, a carrier ampholyte-generated pH gradient is in reality a multi step gradient where each step is made by a different chemical species, and the height of the step is the concentration of this species. Needless to say, it is impossible to reach such a degree of control during the syntheses of different ampholyte batches, so that local deformations are unavoidable. In spite of all these problems, 2D gel electrophoresis was already able to separate over 1500 protein spots, as shown in Figure 1, a performance close to what is achievable today.

So the future looked bright. But besides this nice evolution in protein biochemistry a real revolution was taking place at the same time elsewhere in molecular biosciences, in the nucleic acid world, with cloning of DNA [15], including cDNAs [16], and DNA sequencing [17, 18].

2. Ad augusta per angusta (to brightness through darkness): the 80's, from underdog days to the birth of proteomics

At the very beginning of the 80's, the situation in the microanalysis of the major cellular macromolecules had toppled. Long before PCR was invented, DNA cloning was a way to amplify DNA up to the amounts that were needed to read a sequence, and basically any gene and any mRNA, thanks to the cDNA trick, could be fully deciphered. In addition, cloning made heterologous expression possible [19], so that only sky was the limit for genetic engineering.

In contrast, as proteins could not (and still cannot) be amplified directly without resorting to nucleic acids, 2D electrophoretic maps were basically mute. Of course antibodies were already a very powerful identification tool [20], but the huge gap between the abilities of 2D electrophoresis in terms of protein separation on the one hand, and the requirements of protein sequencing on the other hand, did not allow the protein scientists to answer the simple question: what is the protein that I see in this wonderfully changing spot ?

Thus, 2D maps were very much looking as astronomical star maps, and it is not by chance that one of the first softwares used for the comparison of 2D gels was named from the Renaissance astronomer Tycho Brahe [21]. As the french philosopher Blaise Pascal wrote, "le silence éternel de ces espaces infinis m'effraie" (the eternal silence of the infinite spaces frightens me). So by one of these pendulum swings occurring from time to time in science, many scientists were afraid and switched from the monkish and silent world of proteins to the Babel of DNA, where every piece of DNA could produce a story (and a paper), including some stories of mutations, molecular evolution, orthologs and paralogs, and so on and so forth.

In this context, the community of scientists still believing that large scale analysis of proteins had a future was holding by its fingernails. It was common practice to hear that protein analysis was a thing of the past, and that all what was needed for proteins was biotechnological stuff to purify correctly the recombinant proteins. "I clone therefore I am" was the motto of these days. Thus, resource allocation was scarce for protein scientists, and it is probably not unfair to say that proteomics was close to extinction before it was even born, despite the creativity of the protein scientists of this time. As naming the proteins on 2D gels was not possible, the 2D gellers had developed computerized gel analysis systems [22-24] pattern analysis and global analysis tools to derive biological sense from their maps [25-29], doing profiling long before anyone else. Thus, it is a missense to say that modern computerized analysis of large datasets has been pioneered by transcriptomics. Once again, as in sequencing, proteins were first and nucleic acids second. However, despite all these fancy tools, a robust protein identification method applicable to 2D gel spots was needed if the soon-to-come proteomics was to survive. Ten years after the glorious start in 1975, it was really the bottom of the tide. Hopefully, two major improvements were developing almost underground in this decade of triumphant DNA. The first major improvement was the development of immobilized pH gradient, in other words a chemically clean way to produce high - performance and tailorable pH gradients. It took half a decade to Bjellqvist, Righetti and Gorg, the Holy Trinity of immobilized pH gradients, to develop this promising technique [30] into a really usable tool (reviewed in [31]). Many problems were first to be solved to enable just isoelectric focusing of most proteins, from gel polymerization [32] to pH gradient design [33-35] and even fighting strong adsorption problems [36-38]. Then producing a really practicable solution to interface immobilized pH gradients with SDS gels was not easy either [39] and required quite extensive work [40-43]. However, by the end of the decade, 2D electrophoresis was stronger than ever, thanks to immobilized pH gradients, with high reproducibility [44], high micropreparative abilities [45, 46], further increased later by improved protein extraction solutions [47], and finally high sensitivity of detection without radioisotopes thanks to silver staining [48], which mechanisms were finally understood [49] and controlled after a decade of collective work started in 1979 [50], going here again from black magics to well-controlled laboratory routine.

Although important, these developments were not crucial. The key development was at the protein identification stage, and the key enabling technology was the gas phase sequencer [51]. On the one hand it decreased the amount of protein needed from the nanomole to the low picomole range, and on the other hand the interface between 2D gels and the sequencer was much easier than ever before. Blotting on PVDF membranes allowed the direct sequencing of N-terminally free proteins [52], while techniques were developed to digest the proteins, separate the peptides and sequence them, either from blots [53-55] or from gels [56], developing the in-gel digestion process that is still widely in use today.

When combining these two processes, proteomics became eventually possible. Complex protein samples could be separated by high-resolution 2D electrophoresis, quantified and compared by computerized systems. Then lists of spots of interest could be drawn, and the molecular identity of these spots could be determined by Edman sequencing, sometimes after years of previously unsuccessful effort [57-59]. If the sequence of the gene was available then the story was much easier, but quite often it was not the situation at that time, and the protein was a "novel" one. However, the internal sequences obtained by the Edman process were information-rich, so that it was still possible to derive DNA probes from these peptides, to screen DNA libraries and finally identify the gene of interest [60, 61].
Even for the largest and most performing laboratories in this area, throughputs were quite low, in the order of 1 protein per day. This did not prevent the paleoproteomists to produce a lot of useful data (e.g. in [62-70] ), paving the way for modern proteomics.

3. From Edman sequencing to mass spectrometry: a classical ménage à trois

If we describe 2D gel electrophoresis as the father of proteomics and Edman sequencing as its mother, then the switch from Edman sequencing to mass spectrometry that occurred soon after the birth of proteomics is some kind of middle age lust crisis, i.e. switching from an old wife to a younger, more attractive one.
In 1981, when Edman sequencing gained its full maturity with the gas phase sequencer [51], mass spectrometry of macromolecules was still in its infancy, with FAB ionization just coming out [71]. This was however the first method able to analyze large and polar molecules such as peptides [72], but half a decade was needed for more practical methods to be applied for peptide ionization [73-75]. So at the end of the 80's both mass spectrometry and Edman sequencing had around the same performances in sensitivity and in throughput, with the incredible advantage for mass spectrometry of being able to analyze modified amino acids [76, 77] even with non classical modifications [78]. However, reading a sequence by mass spectrometry required interpreting manually the MS/MS fragmentation spectra [79, 80] a process that required as much expertise as running an Edman sequencer. On such bases, and as science is normally based on performance rather than on fashion, how did it come that mass spectrometry overwhelmed Edman sequencing in a few years? Well, as in every good vaudeville, there is a lover hidden in the closet. Ironically enough, this lover is the old archenemy that almost killed proteomics in its tracks in the 80's, namely DNA sequencing.
The fact is that Edman sequencing produces high quality sequence data, with almost no holes, but this quality comes at the price of low productivity. These data are ideal when the gene sequence is not necessarily known, but this level of quality is somehow superfluous when the name of the game is just to get an univocal identification in a complete DNA database. Conversely, getting complete peptide sequences with mass spectrometry is also material and labor intensive, but mass spectrometry can generate easily and with a high productivity data that are slightly fragmentary, but that are still valuable enough to produce an univocal identification in databases. Furthermore, tricks can be developed that allow to perform automated searches against databases. Maybe the nicest of these tricks is peptide mass fingerprinting [81-84], which allows a univocal identification of a purified protein with no sequence data and only peptide masses, provided that the full length cDNA sequence is available. With the incredible progresses in DNA sequences [85, 86], hallmarked by the first completely sequenced genomes in 1995 [87], this trick became efficient enough to reach the goal of protein identification, with a tremendous increase in productivity. Soon after came the computerized search of MS/MS data [88, 89], and with more data (partial amino acid sequences) came the ability to analyze mixtures and not only purified proteins [90].
With this quantum leap in protein quantification came proteomic strategies based on less demanding analyte separation techniques than 2D electrophoresis, as in the various flavors of shotgun proteomics [91, 92]. The rest is well-known recent history, and other authors in this issue will give a much better view of the proteomic age, as opposed to the paleoproteomic period.

4. Coda

The rather hectic development of proteomics, as related here, reflects the fact that proteomics has always been, and still is, a lame science. Lame by the fact that its two legs, namely separation science and identification, have never had the same length. It started by separation without identification, but now the reverse is true, and identification is well beyond separation. In some kind of parthenogenetic fever, there is a current trend in proteomics stating that mass spectrometry will do almost everything with minimal separation prior to it. However, it has been shown over and over that good separation dramatically improves the coverage of the proteome [93, 94]. Proteomics appears rather short of new and high performance separation methods, as two-dimensional separation of peptides was described almost 30 years ago [95] and has just been miniaturized. Clearly enough, existing separation methods are not able to take the challenges of the number of analytes and dynamic expression range, and they are the factor limiting the performance of proteomics, and especially its reproducibility [96]. Within this frame, it is also my personal opinion that isoelectric focusing has still a lot to offer in modern proteomics, as its latest flavors have not been fully exploited. For example, segmented IEF with isoelectric membranes [97, 98], if applied to peptide IEF, would bring a reproducibility that is lacking and that should dramatically improve the overall reproducibility of shotgun proteomics and allow further tailoring of the separation prior to the MS analysis.

It is also my old timer opinion that moderation in all things is wise. Thus, proteomics should not indulge too much neither in fashion arguments nor in the general recent trend in omics that prizes quantity over quality. DNA sequencing is not free of this problem either, as now more time is spent and more errors are made in assembling genomes than in sequencing their DNA.

Moreover, in the constant competition between proteomics and transcriptomics, basic chemical diversity reasons in peptides compared to polynucleotides will always make proteomics more difficult than transcriptomics. Thus, in a side to side race, proteomics will always lose. Consequently, proteomics should concentrate more on its unique strengths, namely the analysis of all that happens only in the protein world, independently of RNA (basically PTMs and protein complexes). With complex genomes that are not that complex in terms of protein coding genes (only 10% more in man compared to C. elegans) it is obvious that a lot of regulations that make living organisms so complex lay at the protein level, awaiting for next generation proteomics to be discovered. Here again, it will require a countercurrent switch in paradigms, from quantity back to quality, from brute force back to subtlety. It is time to leave the safe shores of boolean logics and to go into the more inhospitable areas of fuzzy logics, where things can be both Yin and Yang at the same time. We are still using the classical molecular genetics paradigm, where correctness of sequence and level of expression is everything. We shall move to a more subtle paradigm, where the quality or even the "flavor" of proteins, as defined by how they are modified, reversibly or not, will be a key element as well. It is only a problem of will, as proteomics has already shown that it could take this challenge [99-101]. Furthermore, the recent developments of top-down proteomics hold great promises in this direction [102]. Mass spectrometry offers the ability to decipher any type of modification, an opportunity that has never been encountered before in biosciences, and it would be pity to keep blinkered and to analyze only peptides as they are predicted from DNA databases.

It is true that proteomics, as all omics, is a technology-driven science, and I hope that this paper has shown how difficult the development of proteomics has been. However, the toolbox is now more mature than ever before, and it is now time in this technology-driven science to do less technology and more science, in other words to take more care about the biological sense that proteomics will produce [103]. In science too, a tree is judged by its fruit.

Finally, it should emphasized that the scientific community that gave rise to proteomics was really small, probably no more than 300 scientists all around the world in the 80's, quite a tiny fraction of the molecular biosciences community taken as a whole. As life on earth at the end of Permian the embryonic proteomic community almost disappeared, but it managed to survive, and the whole proteomic community of nowadays owes a lot to this small bunch of pioneers. Just measuring up to these masters and carrying the torch of proteomics through the challenges to come in increasing our understanding of life will be an achievement.
On a more personal note, it has been a privilege to start in this field almost from the very beginning (I started in 1980), and despite the hard times it has been a tremendous reward to see what proteomics has become and a lot of fun and excitement to be part of this play.

validation. Proteomics. 2014;14:157-61

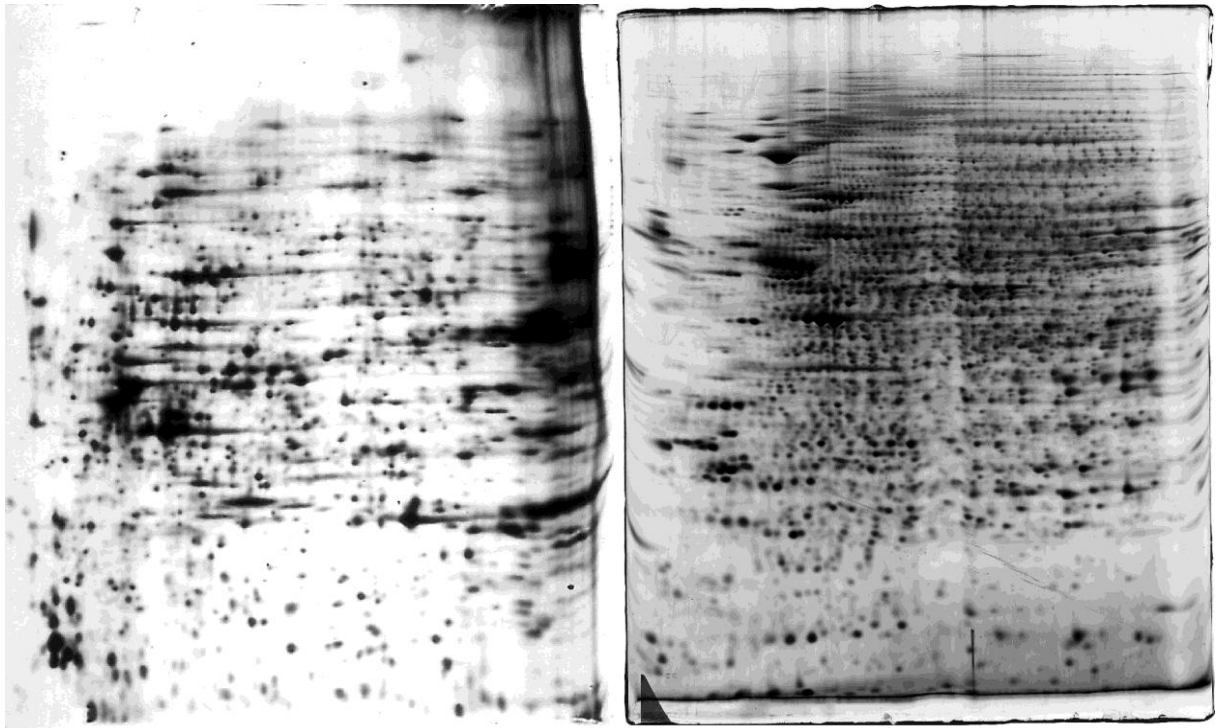

A  B

Figure 1: Evolution of 2D electrophoresis gels over 25 years
Left panel: electrophoresis of total proteins from a human cancer keratinocyte cell line. Carrier ampholytes-driven pH gradient extending from pH 4 to pH 7.2. 35S labelled proteins (with radiolabelled methionine) were used, and 33,000 Bq of radioactive proteins were loaded on the first dimension gel rod. 1304 spots are detectable on the autoradiographic film after 3 weeks of exposure. T. Rabilloud's collection, 1987

Right panel: electrophoresis of total proteins from a mouse premonocyte cell line. Immobilized pH gradient gel in the first dimension, extending from pH 4 to pH 8. 120µg loaded on the first dimension gel strip, detection with silver staining. 1977 spots can be detected. T. Rabilloud's collection, 2012